\documentclass[showpacs,aps,prb,preprint]{revtex4}
\usepackage[english]{babel}
\usepackage{epsfig}
\usepackage{graphicx}
\usepackage{amsmath}
\usepackage{amssymb}

\begin{document}

\title{Phase separation in a two-band model for strongly correlated electrons}

\author{A.~O.~Sboychakov, K.~I.~Kugel, and A.~L.~Rakhmanov}
\affiliation{Institute for Theoretical and Applied
Electrodynamics, Russian Academy of Sciences, Izhorskaya Str. 13,
Moscow, 125412 Russia}

\begin{abstract}
The two-band Hubbard model is used to analyze a possibility of a
non-uniform charge distribution in a strongly correlated electron
system with two types of charge carriers. It is demonstrated that
in the limit of strong on-site Coulomb repulsion, such a system
has a tendency to phase separation into the regions with different
charge densities even in the absence of magnetic or any other
ordering. This tendency is especially pronounced if the ratio of
the bandwidths is large enough. The characteristic size of
inhomogeneities is estimated accounting for the surface energy and
the electrostatic energy related to the charge imbalance.
\end{abstract}



\date{\today}

\maketitle

\section{Introduction}

The phase separation is commonly considered as an inherent
property of strongly correlated electron systems~\cite{DagSci}.
Usually this phenomenon is treated as a result of a coexistence
and competition of different kinds of ordering (magnetic, charge,
orbital)~\cite{dagbook,Kak}. The most widely discussed type of the
phase separation is a formation of nanoscale inhomogeneities such
as ferromagnetic metallic droplets in an insulating
antiferromagnetic material arising due to the self-trapping of
charge carriers~\cite{Nag67}. Such type of the phase separation is
characteristic of the doped manganites. The another type of the
nanoscale inhomogeneity is the modulation of the electron density
due to antiferromagnetic correlations, which is considered as
possible mechanism of the phase separation observed in
superconducting cuprates~\cite{NagUsp95,KivEm}.

Nevertheless, the phase separation can manifest itself even
without some specific type of ordering, e.g., if the system
contains different types of charge carriers. The simplest
illustration of such a behavior gives the Falicov-Kimball
model~\cite{fal}, which is often used as a toy model for
heavy-fermion compounds. This model describes the system of
itinerant and localized electrons with a strong on-site Coulomb
repulsion. The numerical simulations of this model demonstrated an
inhomogeneous charge density distribution at some relation between
the itinerant electron bandwidth and the distance between the
localized level and the bottom of the band~\cite{PSfal}. The
competition between the metallicity and localization in a similar
system with magnetic interactions was studied in
Refs.~\onlinecite{prl,prb} with an emphasis on the phase diagram
of magnetic oxides such as manganites. The system with a band and
localized level is a limiting case of much more common situation
of two bands with different width.

In this paper, we use a two-band Hubbard model for the description
of a strongly correlated electron system with two types of charge
carriers. We demonstrate that the phase separation in this system
arises even without any ordering if the ratio of the bandwidths is
large enough. In Section II, we write out the Hamiltonian of the
model. In Section III, we study the electron structure of a
homogeneous state. In Section IV, we analyze the possibility of
the phase separation and estimate the size of inhomogeneity
accounting for the long-range Coulomb interaction and surface
energy. In Section V, we discuss the obtained results.

\section{The model}

Let us consider a strongly correlated electron system with two
bands $a$ and $b$ of different width. Let the first band, $a$, be
wider than the second one, $b$. Such a system could be described
by the following Hubbard Hamiltonian
\begin{eqnarray}\label{H}
H=-\sum_{\langle\mathbf{ij}\rangle\alpha,\sigma}t^{\alpha}a^{\dag}_{\mathbf{i}\alpha\sigma}a_{\mathbf{j}\alpha\sigma}%
-\epsilon\sum_{\mathbf{i}\sigma}n_{\mathbf{i}b\sigma}
-\mu\sum_{\mathbf{i}\alpha,\sigma}n_{\mathbf{i}\alpha\sigma}
+\frac{1}{2}\sum_{\mathbf{i}\alpha,\sigma}U^{\alpha}n_{\mathbf{i}\alpha\sigma}n_{\mathbf{i}\alpha\bar{\sigma}}%
+\frac{U'}{2}\sum_{\mathbf{i}\alpha,\sigma\sigma'}n_{\mathbf{i}\alpha\sigma}n_{\mathbf{i}\bar{\alpha}\sigma'}\,.
\end{eqnarray} Here, $a^{\dag}_{\mathbf{i}\alpha\sigma}$ and
$a_{\mathbf{i}\alpha\sigma}$ are the creation and annihilation
operators for electrons in the bands $\alpha=\{a,\,b\}$ at site
$\mathbf{i}$ with spin projection $\sigma$, and
$n_{\mathbf{i}\alpha\sigma}=a^{\dag}_{\mathbf{i}\alpha\sigma}a_{\mathbf{i}\alpha\sigma}$.
The symbol $\langle\dots\rangle$ denotes the summation over
nearest-neighbor sites. The first term in Eq.~\eqref{H}
corresponds to the kinetic energy of the conduction electrons in
bands $a$ and $b$ with the hopping integrals $t_a>t_b$. In our
model, we ignore the interband hopping. The second term describes
the shift $\epsilon$ of the center of band $b$ with respect to the
center of band $a$. The last two terms correspond to the on-site
Coulomb repulsion of two electrons either in the same state (with
the Coulomb energy $U^{\alpha}$) or in the different states
($U'$). The bar above $\alpha$ or $\sigma$ denotes {\it not}
$\alpha$ or {\it not} $\sigma$, respectively. The assumption of
the strong electron correlations means that the Coulomb
interaction is large, that is, $U^{\alpha},\,U'\gg
t^{\alpha},\,\epsilon$. The total number $n$ of electrons per site
is a sum of electrons in the $a$ and $b$ states, $n=n_a+n_b$, and
$\mu$ is the chemical potential. Below for definiteness sake, we
consider the case $n\leq 1$.

\section{Homogeneous state}\label{HOM}

The homogeneous state of the model formulated above can be
analyzed by standard methods at arbitrary band filling $n$. Let us
introduce a one-particle Green function
\begin{equation}\label{Grf1}
G_{\alpha\sigma}(\mathbf{j}-\mathbf{j}_0,\,t-t_0)=
-i\langle\hat{T}a_{\mathbf{j}\alpha\sigma}(t)a^{\dag}_{\mathbf{j}_0\alpha\sigma}(t_0)\rangle,
\end{equation}
where $\hat{T}$ is the time ordering operator. The equation of
motion for the one-particle Green function with Hamiltonian
\eqref{H} can be written as
\begin{eqnarray}\label{G}
&&\left(i\frac{\partial}{\partial
t}+\mu+\epsilon^{\alpha}\right)G_{\alpha\sigma}(\mathbf{j}-\mathbf{j}_0,\,t-t_0)=\delta_{\mathbf{jj}_0}\delta(t-t_0)%
-t^{\alpha}\sum_{{\bf\Delta}}G_{\alpha\sigma}(\mathbf{j}-\mathbf{j}_0+\mathbf{\Delta},\,t-t_0)\nonumber\\%
&&+U^{\alpha}{\cal{G}}_{\alpha\sigma,\alpha\bar{\sigma}}(\mathbf{j}-\mathbf{j}_0,\,t-t_0)%
+U'\sum_{\sigma'}{\cal{G}}_{\alpha\sigma,\bar{\alpha}\sigma'}(\mathbf{j}-\mathbf{j}_0,\,t-t_0)\,,
\end{eqnarray}
where $\epsilon^{\alpha}=0$ for $\alpha=a$ and
$\epsilon^{\alpha}=\epsilon$ for $\alpha=b$, the summation in the
second term in the right-hand side of Eq.~\eqref{G} is performed
over sites nearest to $\mathbf{j}$, and $\mathbf{\Delta}$ are the
vectors connecting the site $\mathbf{j}$ with its nearest
neighbors. Equation \eqref{Grf1} includes the two-particle Green
functions of the form
\begin{equation}\label{F}
{\cal{G}}_{\alpha\sigma,\beta\sigma'}(\mathbf{j}-\mathbf{j}_0,\,t-t_0)=%
-i\langle\hat{T}a_{\mathbf{j}\alpha\sigma}(t)n_{\mathbf{j}\beta\sigma'}(t)%
a^{\dag}_{\mathbf{j}_0\alpha\sigma}(t_0)\rangle\,.
\end{equation}
Then, we should write the equations of motion for these functions,
which will include the next order Green functions, etc. To cut
such infinite chain of equations, we shall use here the following
procedure.

In the limit of strong Coulomb repulsion, the presence of two
electrons at the same site is unfavorable, and the two-particle
Green function Eq.~\eqref{F} is of the order of $1/U$, where
$U\sim U_{\alpha},U'$. The equation of motion for
${\cal{G}}_{\alpha\sigma,\beta\sigma'}$ includes the
three-particle terms coming from the commutator of
$a_{\mathbf{j}\alpha\sigma}(t)$ with the $U$ terms of
Hamiltonian~\eqref{H} in the form $\langle\hat{T}a_{\mathbf{j}\alpha\sigma}(t)n_{\mathbf{j}\beta\sigma'}(t)%
n_{\mathbf{j}\gamma\sigma''}(t)a^{\dag}_{\mathbf{j}_0\alpha\sigma}(t_0)\rangle$.
It is easy to see that these terms are of the order of $1/U^2$ and
in our approximation, we neglect them. In the equations of motion
for the two-particle Green functions, we make the decoupling
corresponding to the Hubbard I approximation~\cite{Hub}. That is,
in term coming from the commutator of
$a_{\mathbf{j}\alpha\sigma}(t)$ with the kinetic-energy terms of
Hamiltonian~\eqref{H}, we make the following replacement
$\langle\hat{T}a_{\mathbf{j}+\mathbf{\Delta}\alpha\sigma}(t)n_{\mathbf{j}\beta\sigma'}(t)%
a^{\dag}_{\mathbf{j}_0\alpha\sigma}(t_0)\rangle\to\langle%
n_{\mathbf{j}\beta\sigma'}\rangle\langle\hat{T}%
a_{\mathbf{j}+\mathbf{\Delta}\alpha\sigma}(t)a^{\dag}_{\mathbf{j}_0\alpha\sigma}(t_0)\rangle$.
The analogous decoupling in terms coming from the commutator of
$n_{\mathbf{j}\alpha\sigma}(t)$ with the same kinetic-energy
operator yields zero~\cite{prl,prb,Hub}. As a result, the
equations for the two-particle Green functions can be written as
\begin{eqnarray}
&\left(i\displaystyle\frac{\partial}{\partial
t}+\mu+\epsilon^{\alpha}-U^{\alpha}\right)&{\cal{G}}_{\alpha\sigma,\alpha\bar{\sigma}}(\mathbf{j}-\mathbf{j}_0,\,t-t_0)\nonumber\\%
&&=n_{\alpha\bar{\sigma}}\left[\delta_{\mathbf{jj}_0}\delta(t-t_0)%
-t^{\alpha}\sum_{{\bf\Delta}}G_{\alpha\sigma}(\mathbf{j}-\mathbf{j}_0+\mathbf{\Delta},\,t-t_0)\right]\label{F1}\\%
&\left(i\displaystyle\frac{\partial}{\partial
t}+\mu+\epsilon^{\alpha}-U'\right)&{\cal{G}}_{\alpha\sigma,\bar{\alpha}\sigma}(\mathbf{j}-\mathbf{j}_0,\,t-t_0)\nonumber\\%
&&=n_{\bar{\alpha}\sigma}\left[\delta_{\mathbf{jj}_0}\delta(t-t_0)%
-t^{\alpha}\sum_{{\bf\Delta}}G_{\alpha\sigma}(\mathbf{j}-\mathbf{j}_0+\mathbf{\Delta},\,t-t_0)\right]\label{F2},%
\end{eqnarray}
where $n_{\alpha\sigma}=\langle n_{\mathbf{j}\alpha\sigma}\rangle$
is the average number of electron per site in the state
$(\alpha,\sigma)$.

Eqs.~\eqref{G},~\eqref{F1}, and~\eqref{F2} are the closed system
for one- and two-particle Green functions. This system can be
solved in a conventional manner~\cite{Hub} passing from the
time-space to the frequency-momentum representation. Eliminating
two-particle Green functions we can find the explicit expression
for single-particle Green functions, which have poles
corresponding to the Hubbard sub-bands for each band, $a$ or $b$.
The splitting between these sub-bands is determined by on-site
Coulomb repulsion $U^{\alpha}$ of electrons in the same states.
The on-site Coulomb repulsion $U'$ of electrons in different
states gives rise to the correlation between the fillings of $a$
and $b$ bands (analogous to the case of localized and itinerant
electrons discussed in Ref.~\onlinecite{prl,prb}).

If the total number of the electrons per site does not exceed
unity, $n\leq1$, the upper Hubbard sub-bands are empty, and we can
proceed to the limit $U_{\alpha},\,U'\rightarrow\infty$. In this
case, the one-particle Green function $G_{\alpha\sigma}$ in the
frequency-momentum representation can be written as
\begin{equation}\label{Ginf}
G_{\alpha\sigma}(\mathbf{k},\omega)=
\frac{g_{\alpha\sigma}}{\omega+\mu+\epsilon^{\alpha}-g_{\alpha\sigma}w_{\alpha}\zeta(\mathbf{k})}\,,
\end{equation}
where $w_{\alpha}=zt^{\alpha}$, $z$ is the number of nearest
neighbors,
\begin{equation}\label{g}
g_{\alpha\sigma}=1-\sum_{\sigma'}n_{\bar{\alpha}\sigma'}-n_{\alpha\bar{\sigma}}\,,
\end{equation}
and
$$
\zeta(\mathbf{k})=-\frac{1}{z}\sum_{\mathbf{\Delta}}\text{e}^{i\mathbf{k\Delta}}
$$
is the spectral function depending on the lattice symmetry. In the
case of simple cubic lattice, we have
\begin{equation}\label{zeta}
\zeta({\bf k})=-\frac{1}{3}\left[\cos (k^1d)
+\cos(k^2d)+\cos(k^3d)\right],
\end{equation}
where $d$ is the lattice constant.

In the main approximation in $1/U$ considered here, the magnetic
ordering does not appear. To study the possible types of magnetic
ordering in our model, it is necessary to take into account the
higher order terms in $1/U$. So, below we assume that
\begin{equation}\label{pm}
 n_{\alpha\uparrow}=n_{\alpha\downarrow}\equiv n_{\alpha}/2.
\end{equation}
From the expression for Green function \eqref{Ginf}, it follows
that the filling of each lower sub-band is equal to
$g_{\alpha\uparrow}+g_{\alpha\downarrow}\equiv 2g_\alpha$. Using
\eqref{g} and \eqref{pm}, we have
\begin{equation}\label{gp}
g_{\alpha}=1-n_{\bar{\alpha}}-\frac{n_{\alpha}}{2}\,.
\end{equation}

Using the relation for the density of states
$\rho_{\alpha}(E)=-\pi^{-1} \textmd{Im}\int
G_{\alpha}(\mathbf{k},E+i0)d^3{\bf k}/(2\pi)^3$, we get the
following expression for the number of electrons
\begin{equation}\label{nalpha}
n_{\alpha}=2g_{\alpha}n_0\left(\frac{\mu+\epsilon^{\alpha}}{g_{\alpha}w_{\alpha}}\right)\,
\end{equation}
where
\begin{equation}\label{n0}
n_0(\mu')=\int\limits_{-1}^{\mu'}dE'\,\rho_0(E')\,,
\end{equation}
and
\begin{equation}\label{rho0}
\rho_0(E')=\displaystyle\int\frac{d^3\mathbf{k}}{(2\pi)^3}\,\delta(E'-\zeta(\mathbf{k}))
\end{equation}
is the density of states for free electrons (with the energy
normalized by unity, $|E|\leq1$). The chemical potential $\mu$ in
Eq.~\eqref{nalpha} can be found from the equality $n=n_a+n_b$.

\begin{figure}
\begin{center}
\epsfig{file=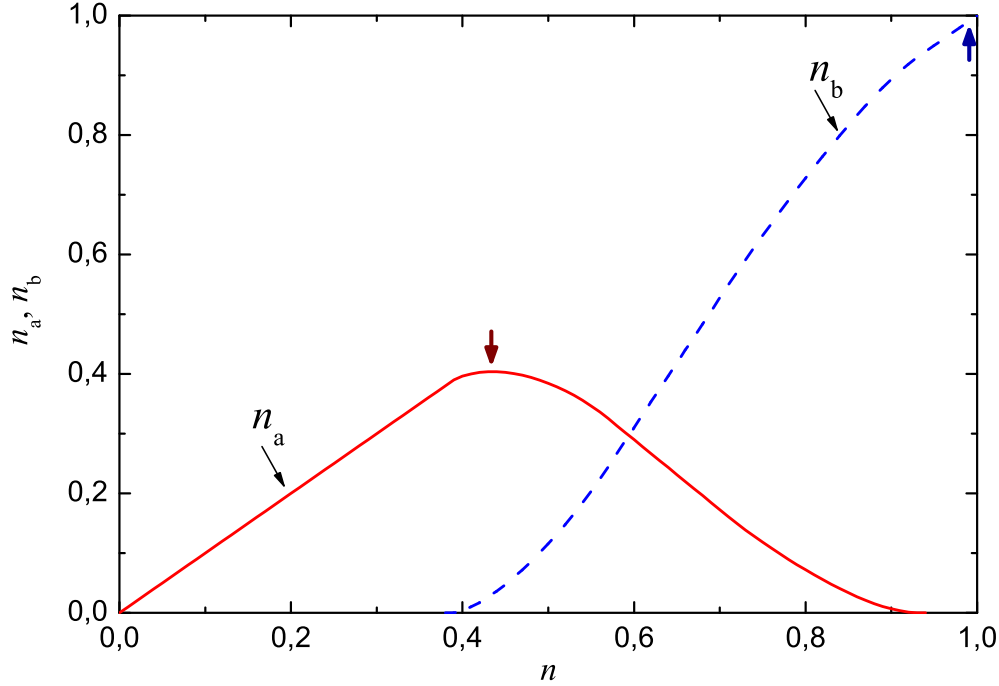,width=0.8\textwidth}
\end{center} \caption{\label{FigN} (Color online) $n_a$ (solid line) and $n_b$
(dashed line) \textit{vs} the total number of charge carriers $n$;
$w_b/w_a=0.2$, $\epsilon/w_a=0.12$. Vertical arrows show the
concentrations of $n_a$ and $n_b$ in the inhomogeneous state (see
the text below).}
\end{figure}

\begin{figure}
\begin{center}
\epsfig{file=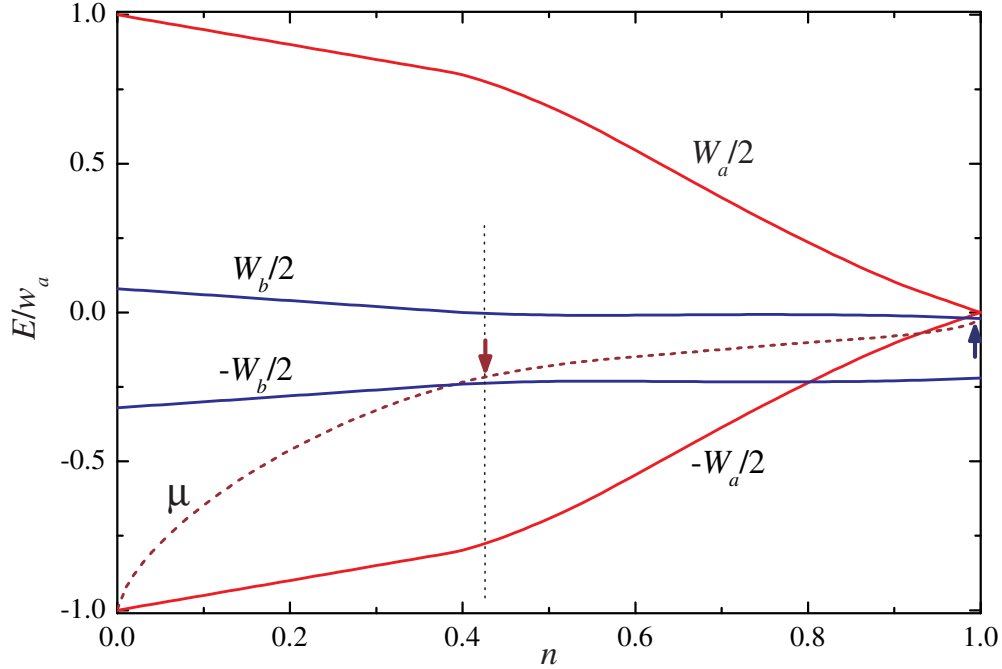,width=0.8\textwidth}
\end{center}
\caption{\label{FigW} (Color online) Effective bandwidths
$W_{\alpha}=2w_ag_{\alpha}$ \textit{vs} the total number of charge
carriers $n$. The dashed curve is the chemical potential $\mu$.
The values of the parameters are $w_b/w_a=0.2$,
$\epsilon/w_a=0.12$. Vertical arrows show the concentrations of
$n_a$ and $n_b$ in the inhomogeneous state.}
\end{figure}

Let us consider the case when the energy difference $\epsilon$
between centers of $a$ and $b$ bands is not too large, that is, of
the order of the width of $b$ band, $w_b$. In this case, there
exist only $a$ electrons at low doping until the chemical
potential reaches the bottom of the $b$ band $-\epsilon-w_b$ at
some concentration $n_c$. At $n>n_c$, the $b$ electrons appear in
the system, and the effective width of $a$ band,
$W_a=2w_ag_{a}(n_a,n_b)$, starts to decrease. The plots of $n_a$,
$n_b$, and the effective bandwidth are shown in Fig.~\ref{FigN}
and Fig.~\ref{FigW} as functions of $n$. In all calculations, we
use the spectrum $\zeta(\mathbf{k})$ in the form~\eqref{zeta}.

The energy of the system in homogeneous state, $E_{\text{hom}}$,
is the sum of electron energies in all filled bands. Using the the
density of states for free electrons Eq.~\eqref{rho0} we can write
$E_{\text{hom}}$ in the following form
\begin{equation}\label{Ehom}
E_{\text{hom}}=2\sum_{\alpha}g_{\alpha}^2w_{\alpha}%
\varepsilon_0\left(\frac{\mu+\epsilon^{\alpha}}{g_{\alpha}w_{\alpha}}\right)%
-\epsilon n_b\,,
\end{equation}
where
\begin{equation}\label{eps}
\varepsilon_0(\mu')=\int\limits_{-1}^{\mu'}dE'E'\,\rho_0(E')\,.
\end{equation}
The dependence of $E_{\text{hom}}(n)$ is shown in Fig.~\ref{FigE}
by the solid line.

\section{Phase separation}

\subsection{General consideration}

In this section, we analyze the possibility of the phase
separation in the system. As one can see from Fig.~\ref{FigE}, the
energy for the homogeneous state, $E_{\text{hom}}(n)$, has two
minima at different values of the charge carrier density. In this
situation, it is favorable for a system to form two phases with
different electron concentrations. Moreover, the existence of two
minima is not a necessary condition for the formation of an
inhomogeneous state and this phenomenon could be observed under
more general conditions~\cite{prl,prb}. However, the phase
separation could be hindered by increase of the energy due to
surface effects and a charge redistribution.

At first, we do not take into account the charge redistribution
and surface terms in the total energy. In this way, we determine
the optimum content of each phase and the corresponding charge
densities but we can not find the characteristic size of the
inhomogeneities.

We consider two phases, I (low carrier density) and II (high
carrier density), with the number of electrons per site $n_1$ and
$n_2$, respectively. A fraction $p$ of the system volume is
occupied by the phase I and $1-p$ is a fraction of the phase II.
We seek a minimum of the system energy
\begin{equation}\label{PsE0}
E_{\text{ps}}^0(n_1,n_2)=pE_{\text{hom}}(n_1)+(1-p)E_{\text{hom}}(n_2)
\end{equation}
under the condition of the charge carrier conservation
\begin{equation}\label{ChC}
n=pn_1+(1-p)n_2\,.
\end{equation}
The results of calculations of the system energy in the phase
separated state are shown in Fig.~\ref{FigE} for $w_b/w_a=0.2$ and
$\varepsilon/w_a=0.12$ by the dashed curve. From this figure we
see, that the phase separation exists in the range of $n$ where
both types of charge carriers coexist in  the homogeneous state.
The numerical analysis shows that the concentrations of the charge
carriers in each phase, $n_1$ and $n_2$, vary slowly with $n$,
remaining close to certain optimal values for each phase:
$n_1\approx n_a \approx 0.5$ for the state with low carrier
density, whereas $n_2\approx n_b\approx 1$ for the state with high
carrier density. The phase II can be considered as a Mott-Hubbard
insulator since the corresponding lower Hubbard sub-band is almost
completely filled. If $n$ increases from zero to one, the phase
separation may be favorable when the $n$ achieves the value
corresponding to the energy minimum for the homogeneous state. At
this value of $n$ the content of the phase I $p(n)$ starts to
decrease from $p(n)=1$. With the further increase of $n$, $p(n)$
tends to zero at $n\approx n_2$. Therefore, we can conclude that
the system may separate into metallic and insulating phases in a
certain parameter range. An indication to the phase separation is
a negative curvature of the $E_{hom}(n)$ curve at the right side
from the energy minimum, see Fig.~\ref{FigE}.

\begin{figure}
\begin{center}
\epsfig{file=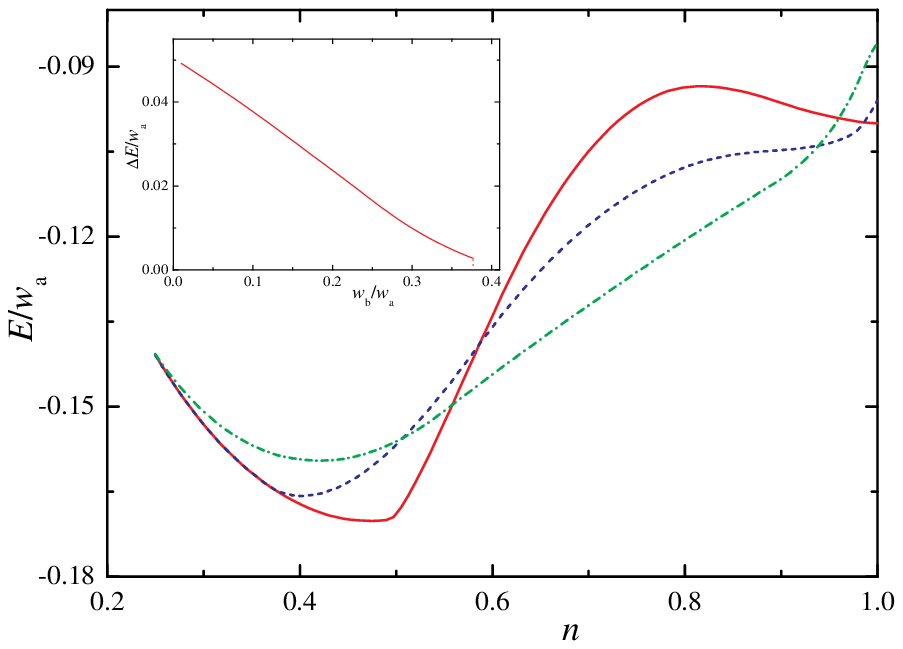,width=0.8\textwidth}
\end{center}
\caption{\label{FigEps} (Color online) The energy of homogeneous
state {\it vs} doping level $n$ at $w_b/w_a=0.1$ (solid line),
$w_b/w_a=0.25$ (dashed line), and $w_b/w_a=0.4$ (dot-dashed line).
The parameter $\varepsilon_b/w_a=0.1$ for all cases. The phase
separation is favorable for solid and dashed curves in the range
of doping $0.45\lesssim n\lesssim1$, where $\partial^2E/\partial
n^2<0$, whereas for dot-dashed curves, only the homogeneous state
exists. In the inset, the maximum energy gain due to the formation
of the inhomogeneous state ($V_0=0$, $\varepsilon_b/w_a=0.1$) as a
function of the ratio $w_b/w_a$ is shown. For $w_b/w_a\gtrsim0.38$
the phase separation becomes unfavorable in energy.}
\end{figure}

The above discussion demonstrates that the width and filling of
one band depends on the width and filling of other bands. The
phase separation gives the possibility to attain the minimum free
energy by an optimum filling of the electron bands. The phase
separation can be favorable only if the bands are appreciably
different. In Fig.~\ref{FigEps} we plot the energy of the
homogeneous state versus $n$ at different values of ratio
$w_b/w_a$. We see that $\partial^2 E/\partial n^2<0$ in a wide
range of $n$ for $w_b/w_a\ll 1$, which indicates the possibility
of the phase separation (see also Fig.~\ref{FigE}). The phase
separation becomes unfavorable only if $w_b/w_a\lesssim 0.4$.

\subsection{Characteristic size of inhomogeneities}

The phase separation leads to redistribution of charge carriers
($n_1\neq n_2$). Therefore, we should take into account the
electrostatic contribution to the total energy of the
phase-separated state. This contribution depends on the shape of
inhomogeneities. Following Refs.~\onlinecite{prl,prb,Lor}, we
consider a spherical geometry of the inhomogeneous state, namely,
at $p<0.5$, the sample is modelled as an aggregate of spheres of
phase I embedded into the matrix of phase II and vice versa for
$p>0.5$. The electrostatic, $E_C$, energy is calculated in the
Wigner-Seitz approximation, that is, we consider a set of
spherical unit cells with zero total charge, where the spherical
core of one phase is surrounded by a shell of another phase.
Following the approach described in Refs.~\onlinecite{prb,Lor} we
find at $p<0.5$
\begin{equation}\label{Ec} E_{\text{C}}=\frac{2\pi e^2}{5\epsilon
d}\left(n_1-n_2\right)^2\left(\frac{R_s}{d}\right)^2p\left(2-3p^{1/3}+p\right),
\end{equation} where $\epsilon$ is the average permittivity and
$R_s$ is the radius of the droplet of the phase I. In the case
$p>0.5$, we should replace $n_1\leftrightarrow n_2$ and
$p\leftrightarrow 1-p$.

\begin{figure}
\begin{center}
\epsfig{file=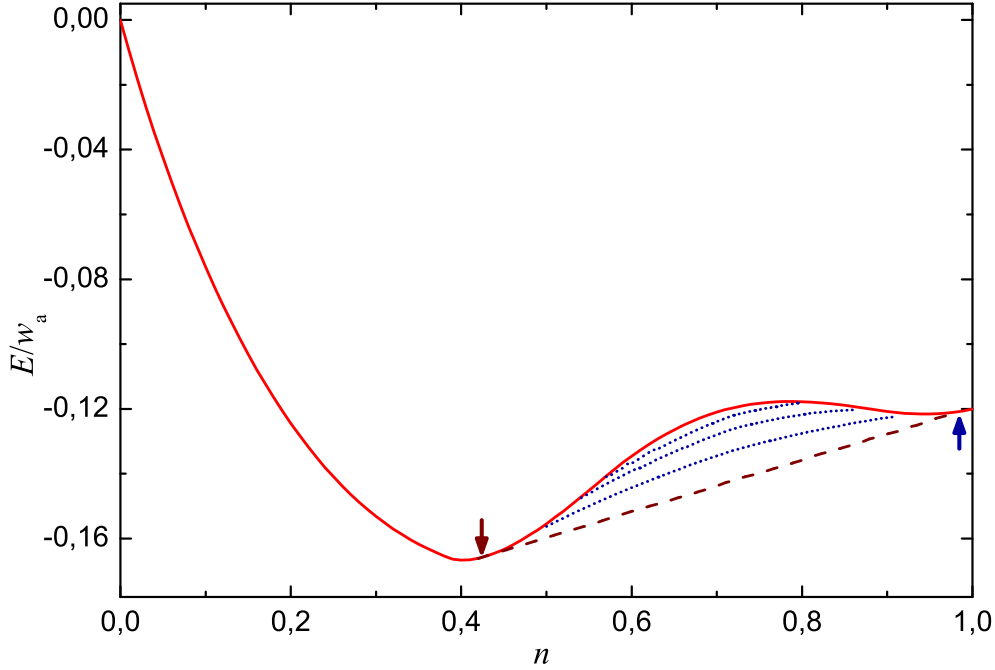,width=0.8\textwidth}
\end{center}
\caption{\label{FigE} (Color online) The energy of the system {\it
vs} doping level $n$. Solid curve corresponds to the homogeneous
state, whereas dashed curve is the energy of phase-separated state
without taking into account electrostatic and surface
contributions to the total energy. Dot curves are the energies of
inhomogeneous state, Eqs.~\eqref{PsE} and \eqref{PsE1}, at
$V_0/w_a=0.1,\,0.05,\,0.01$ from top to bottom (see text below).
Here, $w_b/w_a=0.2$ and $\varepsilon/w_a=0.12$. Vertical arrows
show the concentrations of $n_a$ and $n_b$ in the inhomogeneous
state.}
\end{figure}

Another contribution to the total energy depending of the size of
inhomogeneities is related to the surface between two phases. It
comes from the size quantization and depends on the electron
densities in both phases. The case when one of the densities is
zero was considered in Ref.~\onlinecite{prb}. The generalization
of this approach for non-zero densities is presented in Appendix,
where surface energy $\sigma(n_1,n_2)$ is calculated using the
perturbative approach proposed in Ref.~\onlinecite{BB}. The
corresponding contribution, $E_S$, to the total energy is
proportional to the surface area between phases I and II. At
$p<0.5$, it can be written in the form
\begin{equation}\label{Es}
E_{\text{S}}=p\,\frac{3d}{R_s}\sigma(n_1,n_2)\,,
\end{equation}
In the case $p>0.5$, we should replace $p\to1-p$.

\begin{figure}
\begin{center}
\epsfig{file=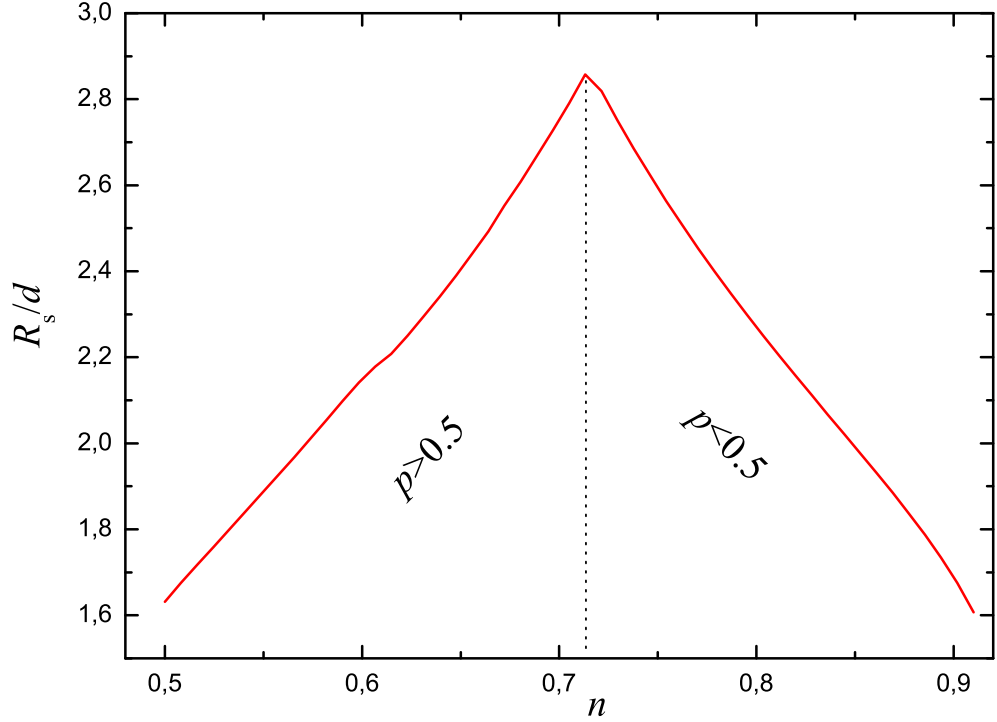,width=0.8\textwidth}
\end{center}
\caption{\label{FigR} (Color online) The radius of droplets $R_s$
{\it vs} doping level $n$ at $w_b/w_a=0.2$, $\epsilon/w_a=0.12$,
and $V_0/w_a=0.02$.}
\end{figure}

Minimization of the sum $E_{\text{CS}}=E_\text{C}+E_\text{S}$ with
respect to $R_s$ allows us to calculate this value. In doing so,
we get at $p<0.5$
\begin{equation}\label{Rs}
R_s=d\left(\frac{15\sigma(n_1,n_2)}{4\pi%
V_0(n_2-n_1)^2\left(2-3p^{1/3}+p\right)}\right)^{1/3}\!\!\!\!\!\!.
\end{equation}
The total energy of the inhomogeneous state then reads
\begin{equation}\label{PsE}
E_{\text{ps}}(n_1,n_2)=pE_{\text{hom}}(n_1)+(1-p)E_{\text{hom}}(n_2)+E_{\text{CS}}(n_1,n_2)\,,
\end{equation}
where
\begin{equation}\label{PsE1}
E_{\text{CS}}=3\left(V_0\frac{9\pi}{10}(n_2-n_1)^2\sigma^2(n_1,n_2)\right)^{1/3}\!\!\!\!\!\!%
p\left(2-3p^{1/3}+p\right)^{1/3}\!\!\!\!\!\!,
\end{equation}
$V_0=e^2/\epsilon d$. The energy $E_{\text{ps}}$ calculated by
minimization of Eq.~\eqref{PsE} with respect to $n_1$ and $n_2$ at
different values of $V_0$ are shown in Fig.~\ref{FigE} by
dot-dashed curves. We see, that the Coulomb repulsion in the
system with an inhomogeneous charge distribution reduces the range
of $n$, in which the phase separation is favorable. In
Fig.~\ref{FigR} we plot the characteristic radius of
inhomogeneities, $R_s$, as a function of $n$.

\section{Discussion}

Thus, the phase separation can be favorable for the system of the
strongly correlated electrons even in the absence of any specific
ordering. We demonstrated that the state with inhomogeneous charge
distribution can arise if there exist two types of the charge
carriers with different bandwidth. The electron correlations due
to on-site Coulomb repulsion lead to the dependence of the
bandwidth for one type of electrons on the band filling for
another type of electrons. As a result, the dependence of energy
on the total number of the charge carriers becomes non-monotonic.
The competition between kinetic and correlation energies triggers
the formation of an inhomogeneous ground state. It is particularly
evident if the energy of the system as a function of electron
density has two minima. In this case, it could be favorable for
the system to separate into two states with electron densities
close to these minima rather than to form a homogeneous state with
an intermediate density. Such a situation is illustrated in
Fig.~\ref{FigE}.

It is clear that the phase separation can occur only if the
bandwidths corresponding to two types of the charge carriers are
sufficiently different, that is, the ratio $w_b/w_a$ of the widths
of narrow and wide band should be rather small. The second
condition is that the narrow and wide bands should not be widely
separated from each other, that is, the ratio $\epsilon/w_a$ of
the distance between the band centers and the width of the wider
band should be less than unity. Naturally, the long-range
electrostatic forces prevent nonuniform charge distribution and
the condition $V_0/w_a\ll 1$ should be met. As it can be seen from
Fig.~\ref{FigEps}, the phase separation can be favorable even if
$w_b/w_a\lesssim 0.3-0.4$.

\section*{Acknowledgments}

The work was supported by the European project CoMePhS,
International Science and Technology Center, grant no. G1335, and
Russian Foundation for Basic Research, project no. 05-02-17600.

\appendix*

\section{Surface energy}\label{SE}

The electrons in the phase-separated state are confined within a
restricted volume $V_s$. This gives rise to the change in the
density of states in both phases. At small ratio
$\Delta=S_sd/V_s$, where $S_s$ is the surface area of the sample,
the density of states for free electrons can be written
as~\cite{prb,BB}
\begin{eqnarray}\label{rhoD2}
\rho(E')=\left(1+\frac{\Delta}{2}\right)\rho_0(E')-\frac{\Delta}{4}
\left(\rho^{(2D)}_0(E'+1/3)+\rho^{(2D)}_0(E'-1/3)\right)\,.
\end{eqnarray}
where $\rho_0$ is given by Eq.~\eqref{rho0}. Here $\rho^{(2D)}_0$
is the density of states in two dimensions. Using this expression
instead of~\eqref{rho0}, and expanding the Eqs.~\eqref{nalpha},
\eqref{n0}, \eqref{Ehom}, and \eqref{eps} in a series in powers of
$\Delta$ up to the first order, we derive formula~\eqref{Es} with
the correction for the size quantization $\sigma(n)$ in the form
\begin{eqnarray}
\sigma=&2&\sum_{\alpha}g_{\alpha}^{(0)}w_{\alpha}%
\left[\varepsilon_0(\mu_{\alpha}^{'(0)})n_{\alpha}^{(1)}+
g_{\alpha}^{(0)}\delta\varepsilon_0(\mu_{\alpha}^{'(0)})\right]+\nonumber\\
&&\sum_{\alpha}g_{\alpha}^{(0)}w_{\alpha}\mu_{\alpha}^{'(0)}\rho_0(\mu_{\alpha}^{'(0)})%
\left(\frac{2\mu^{(1)}}{w_{\alpha}}-\mu_{\alpha}^{'(0)}n_{\alpha}^{(1)}\right)
-\sum_{\alpha}\epsilon n_{b}^{(1)}\,,
\end{eqnarray}
where
\begin{equation}
\mu_{\alpha}^{'(0)}=\frac{\mu^{(0)}+\epsilon^{\alpha}}{g_{\alpha}^{(0)}w_{\alpha}}\,,
\end{equation}
\begin{equation}
n_{\alpha}^{(1)}=\frac12\frac{\displaystyle\frac{4}{w_{\alpha}}\rho_0(\mu_{\alpha}^{'(0)})\mu^{(1)}+g_{\alpha}^{(0)}%
\left[2n_0(\mu_{\alpha}^{'(0)})-n_0^{(2d)}(\mu_{\alpha}^{'(0)}+\frac13)-n_0^{(2d)}(\mu_{\alpha}^{'(0)}-\frac13)\right]}%
{1+\mu_{\alpha}^{'(0)}\rho_0(\mu_{\alpha}^{'(0)})-n_0(\mu_{\alpha}^{'(0)})}\,,
\end{equation}
\begin{equation}
\delta\varepsilon_0(\mu')=\frac12\varepsilon_0(\mu')-\frac14\left[\varepsilon^{(2D)}_0(\mu'+\frac13)+\varepsilon^{(2D)}_0(\mu'-\frac13)\right]%
+\frac{1}{12}\left[n^{(2D)}_0(\mu'+\frac13)-n^{(2D)}_0(\mu'-\frac13)\right],
\end{equation}
correction to the chemical potential $\mu^{(1)}$ is found from the
condition $\sum_{\alpha}n_{\alpha}^{(1)}=0$, and the superscript
$(0)$ denotes the unperturbed value of corresponding quantity. The
functions $n_0^{(2D)}(\mu')$ and $\varepsilon_0^{(2D)}(\mu')$ in
these expressions are determined by Eqs.~\eqref{n0}
and~\eqref{eps}, respectively, where one should change $\rho_0$ to
$\rho_0^{(2D)}$. The surface energy $\sigma(n_1,n_2)$ per unit
area between phases I and II is the sum $\sigma(n_1)+\sigma(n_2)$.

\end{document}